\documentclass[iop,twocolumn]{aastex701}
\usepackage{amsmath}

\hypersetup{
  colorlinks,
  linkcolor={blue!88!black!80},
  citecolor={blue!88!black!80},
  urlcolor={blue!88!black!80}}

\begin{document}

\title{Evidence for a Delayed UV Counterpart to X-ray Quasi-periodic Eruptions in Ansky}

\author[0000-0001-8416-7059]{Hengxiao Guo} 
\affiliation{Shanghai Astronomical Observatory, Chinese Academy of Sciences, 80 Nandan Road, Shanghai 200030, People's Republic of China}
\email[show]{hengxiaoguo@gmail.com (HXG)}  

\author[0000-0002-5385-9586]{Zhen Yan} 
\affiliation{Shanghai Astronomical Observatory, Chinese Academy of Sciences, 80 Nandan Road, Shanghai 200030, People's Republic of China}
\email[show]{zyan@shao.ac.cn (ZY)\\}  

\author[0000-0002-7329-9344]{Ya-Ping Li} 
\affiliation{Shanghai Astronomical Observatory, Chinese Academy of Sciences, 80 Nandan Road, Shanghai 200030, People's Republic of China}
\email[show]{liyp@shao.ac.cn (YPL)\\}

\author[0000-0002-0568-6000]{Joheen Chakraborty} 
\affiliation{MIT Kavli Institute for Astrophysics and Space Research, Massachusetts Institute of Technology, Cambridge, MA 02139, USA}
\email[]{joheen@mit.edu\\}  

\author[0000-0003-0820-4692]{Paula S\'{a}nchez-S\'{a}ez} 
\affiliation{European Southern Observatory, Karl-Schwarzschild-Strasse 2, 85748 Garching bei München, Germany}
\email[]{paula.sanchezsaez@eso.org\\}  

\author[0000-0002-8606-6961]{Lorena Hern\'{a}ndez-Garc\'{i}a} 
\affiliation{Instituto de Estudios Astrof\'isicos, Facultad de Ingenier\'ia y Ciencias, Universidad Diego Portales, Av. Ej\'ercito Libertador 441, Santiago, Chile}
\affiliation{Centro Interdisciplinario de Data Science, Facultad de Ingenier\'ia y Ciencias, Universidad Diego Portales, Av. Ej\'ercito Libertador 441, Santiago, Chile}
\email[]{lorena.hernandez@mail.udp.cl\\}

\author[0000-0003-1702-4917]{Wenda Zhang}
\affiliation{National Astronomical Observatories, Chinese Academy of Sciences, 20A Datun Road, Beijing 100101, People's Republic of China}
\email[]{wdzhang@nao.cas.cn}

\author[0000-0001-8416-7059]{Jingbo Sun} 
\affiliation{Shanghai Astronomical Observatory, Chinese Academy of Sciences, 80 Nandan Road, Shanghai 200030, People's Republic of China}
\email[]{sunjingbo@shao.ac.cn\\}

\author[0000-0002-7299-4513]{Shuang-Liang Li}
\affiliation{Shanghai Astronomical Observatory, Chinese Academy of Sciences, 80 Nandan Road, Shanghai 200030, People's Republic of China}
\email[]{lisl@shao.ac.cn}

\author[0000-0001-6858-1006]{Hongping Deng}
\affiliation{Shanghai Astronomical Observatory, Chinese Academy of Sciences, 80 Nandan Road, Shanghai 200030, People's Republic of China}
\email[]{hpdeng353@shao.ac.cn}

\author[0000-0002-4521-6281]{Wenwen Zuo}
\affiliation{Shanghai Astronomical Observatory, Chinese Academy of Sciences, 80 Nandan Road, Shanghai 200030, People's Republic of China}
\email[]{wenwenzuo@shao.ac.cn}

\author[0000-0002-5674-0644]{Hiromichi Tagawa} 
\affiliation{Shanghai Astronomical Observatory, Chinese Academy of Sciences, 80 Nandan Road, Shanghai 200030, People's Republic of China}
\email[]{htagawa@shao.ac.cn}

\author[0000-0002-6938-3594]{Xin Pan} 
\affiliation{School of Physical Science and Technology, Southwest Jiaotong University, Chengdu 610031, People's Republic of China}
\email[]{panxin@swjtu.edu.cn}

\author[0009-0006-5706-0364]{Minghao Zhang}
\affiliation{National Astronomical Observatories, Chinese Academy of Sciences, 20A Datun Road, Beijing 100101, People's Republic of China}
\email[]{zhangmh@bao.ac.cn}

\author[0000-0001-5675-6323]{Patricia Ar\'{e}valo} 
\affiliation{Millennium Nucleus on Transversal Research and Technology to Explore Supermassive Black Holes (TITANS), Gran Breta\~{n}a 1111, Playa Ancha, Valpara\'{i}so, Chile}
\email[]{patricia.arevalo@uv.cl\\}

\author[0000-0003-1523-9164]{Paulina Lira}
\affiliation{Departamento de Astronom \`{\i}a, Universidad de Chile, Casilla 36D, Santiago, Chile}
\email[]{plira@das.uchile.cl\\}

\author[0000-0002-2006-1615]{Chichuan Jin}
\affiliation{National Astronomical Observatories, Chinese Academy of Sciences, 20A Datun Road, Beijing 100101, People's Republic of China}
\email[]{ccjin@nao.cas.cn}

\author[0000-0002-4455-6946]{Minfeng Gu} 
\affiliation{Shanghai Astronomical Observatory, Chinese Academy of Sciences, 80 Nandan Road, Shanghai 200030, People's Republic of China}
\email[]{gumf@shao.ac.cn\\}

\begin{abstract}

X-ray quasi-periodic eruptions (QPEs) represent a novel population of extreme, repeating nuclear transients whose physical origins remain debated. A defining characteristic of QPEs has been their exclusive detection in the X-ray band, with a notable absence of correlated multi-wavelength counterparts. Here we report the first detection of a recurrent UV response temporally coupled to the X-ray QPE signal in the source Ansky/ZTF19acnskyy. The UV emission displays coherent periodic modulations over five consecutive cycles, systematically lagging the X-ray eruptions by $0.96^{+0.38}_{-0.39}$ days, with a cross-correlation coefficient of $r_{\rm max} \sim 0.6$. We suggest that the detectability of this corresponding signal may be enabled by Ansky's unusually long recurrence timescale, which could reduce the temporal smearing of the UV response seen in more rapid QPEs. The observed delay may correspond to a diffusion timescale associated with heated blobs. However, we cannot exclude the possibility that the lag corresponds to the light-crossing time associated with X-ray irradiation that originates near the central black hole and propagates to the outer UV-emitting region. While numerous QPE models have been proposed, any viable model for Ansky must be able to simultaneously explain the presence of a UV counterpart, its measured time lag, and the previously observed steadily increasing recurrence period.

\end{abstract}




\section{Introduction} 
X-ray quasi-periodic eruptions (QPEs) are a recently discovered class of extreme, repeating nuclear outbursts from accreting black holes \citep{Miniutti19,Arcodia21}. To date, more than ten QPE sources have been identified, including GSN~069 \citep{Miniutti19}, RX~J1301\citep{Sun13,Giustini20}, eRO-QPE1--5 \citep{Arcodia21,Arcodia24a,Arcodia24b,Arcodia25}, XMM J0249 \citep{Chakraborty21}, AT2019qiz \citep{Nicholl24}, AT2022upj \citep{Chakraborty25}, AT2019vcb \citep{Quintin23,Bykov25}, Ansky \citep{Sanchez-Saez24,Hernandez-Garcia25a,Hernandez-Garcia25b}, and eRASSt J2344 \citep{Baldini26}. These systems typically exhibit ultrasoft X-ray spectra, with blackbody temperatures rising from a stable quiescent level of $\sim50$~eV to $\gtrsim100$~eV during eruptions, reaching peak luminosities of $\sim10^{42}$--$10^{44}\ \mathrm{erg\ s^{-1}}$. The eruptions recur on timescales from hours to more than 10 days, with individual burst durations spanning $\sim0.5$~hr to $\sim1.5$~days, and are exclusively found in the nuclei of nearby low-mass galaxies \citep{Miniutti19,Arcodia21,Arcodia25}. The central supermassive black holes (SMBHs) typically have masses of $10^{5}$--$10^{7.5}\,M_\odot$. Their host galaxies show a strong over representation of post-starburst systems and generally lack broad optical emission lines \citep{Wevers22}. Growing observational evidence further suggests that QPEs preferentially arise during the decay phase of tidal disruption events (TDEs), supporting a physical connection between the two phenomena \citep{Chakraborty21,Miniutti23a,Quintin23,Nicholl24,Chakraborty25,Jiang25}.

The physical origin of QPEs remains under intense debate, with three main classes of models currently proposed. One invokes \textit{accretion-disk instabilities}, including thermal--viscous cycles, radiation-pressure instabilities, or disk tearing and precession \citep{Raj21,Pan22,Pan23,Kaur23,Zhang25}. A second class attributes QPEs to \textit{periodic mass transfer} onto the SMBH from one or more orbiting bodies, for example via Roche-lobe overflow in eccentric orbits \citep{King22,Zhao22,Metzger22,Wang22,Krolik22,Linial23a,Lu23}. A third scenario involves \textit{disk--star or disk--stellar-mass black hole collisions}, in which repeated impacts generate shock-heated hot spots that power the observed X-ray flares \citep{Linial23b,Xian21,Tagawa23,Zhou24a,Zhou24b,Huang25,Jankovivc26,Suzuguchi26}

Ansky/ZTF19acnskyy, located in SDSS J133519.91$+$072807.4 at $z=0.024$, is the most extreme member of the QPE population discovered to date, exhibiting the longest recurrence intervals and burst durations observed in any known system \citep{Hernandez-Garcia25a,Hernandez-Garcia25b}. It is also the most luminous QPE source identified so far, reaching peak soft X-ray luminosities of $L_{\rm peak} \sim 4\times10^{43}\ \mathrm{erg\ s^{-1}}$. The source was initially identified as a nuclear transient beginning in late 2019 \citep{Sanchez-Saez24}. Its X-ray emission was first detected a few years after the optical flare. Late-time \textit{HST} UV spectroscopy revealed a featureless steep continuum consistent with a featureless TDE \citep{Zhu25}. Its black hole mass is estimated to be $\log (M_{\rm BH}/M_{\odot}) \sim6-7$ based on various indirect methods, including empirical scaling relations \citep{Sanchez-Saez24,Hernandez-Garcia25a}. Ansky exhibits a QPE state in early 2024 with an initial recurrence period of $\sim4.5$ days and burst durations (defined as the FWHM of each eruption) of $\sim0.6$ days. Since then, it has undergone a dramatic temporal evolution: high-cadence \textit{NICER} monitoring in 2025 revealed a doubling of these timescales, with recurrence periods extending to $\sim10$ days and durations to $\sim1.5$ days, accompanied by a fourfold increase in radiated energy per flare \citep{Hernandez-Garcia25b}. Continued monitoring through February 2026 indicates that this lengthening trend persists, with a period derivative of $\sim 1.7\times10^{-2}$ day per day \citep{Chakraborty26} and a current recurrence period of $\sim14$ days. 

Most importantly, Ansky exhibits reliable UV/optical variability \citep{Sanchez-Saez24,Hernandez-Garcia25a,Zhu25}, distinguishing it from all other known QPE sources, in which variability is typically confined to soft X-rays. Systematic monitoring of Ansky with \textit{Swift}/UVOT revealed year-long UV variability detected at $6\sigma$ in the UVW2 band and $4\sigma$ in the UVM2 band. In addition, simultaneous \textit{XMM-Newton}/OM observations captured rapid UV fluctuations in the UVM2 band at the $3\sigma$ level. These findings motivated a more intensive, high-cadence \textit{Swift} monitoring campaign in late 2025 to investigate the detailed structure of the UV light curve and its connection to the X-ray eruptions.

In this Letter, we present the results of our high-cadence UV monitoring of Ansky, revealing a corresponding UV response and its physical implications. Section~\ref{sec:obs} describes the observations and data reduction, Section~\ref{sec:res} presents the main results, and Section~\ref{sec:dis} discusses their physical implications.


\begin{figure*}[ht!]
\includegraphics[width=\linewidth]{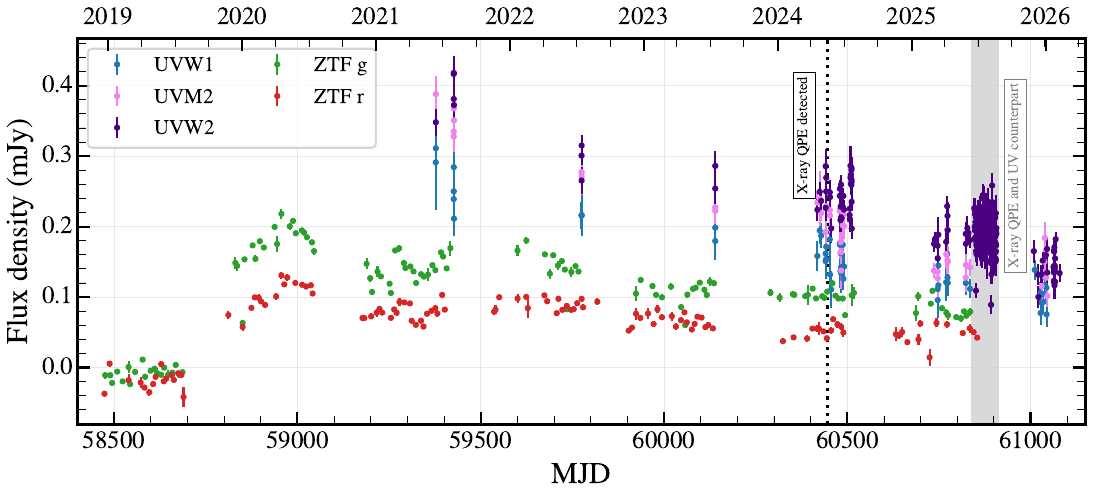}
\vspace{-20pt}
\caption{The long-term ZTF and $Swift$/UVOT light curves after  host subtraction and the Galactic extinction correction. The dotted line marked the time first detected X-ray QPE \citep{Hernandez-Garcia25a}. The gray shadow marked the period when the X-ray QPE and UV counterpart are simultaneously observed.}
\label{fig:longterm}
\end{figure*}

\begin{figure*}[ht!]
\includegraphics[width=\linewidth]{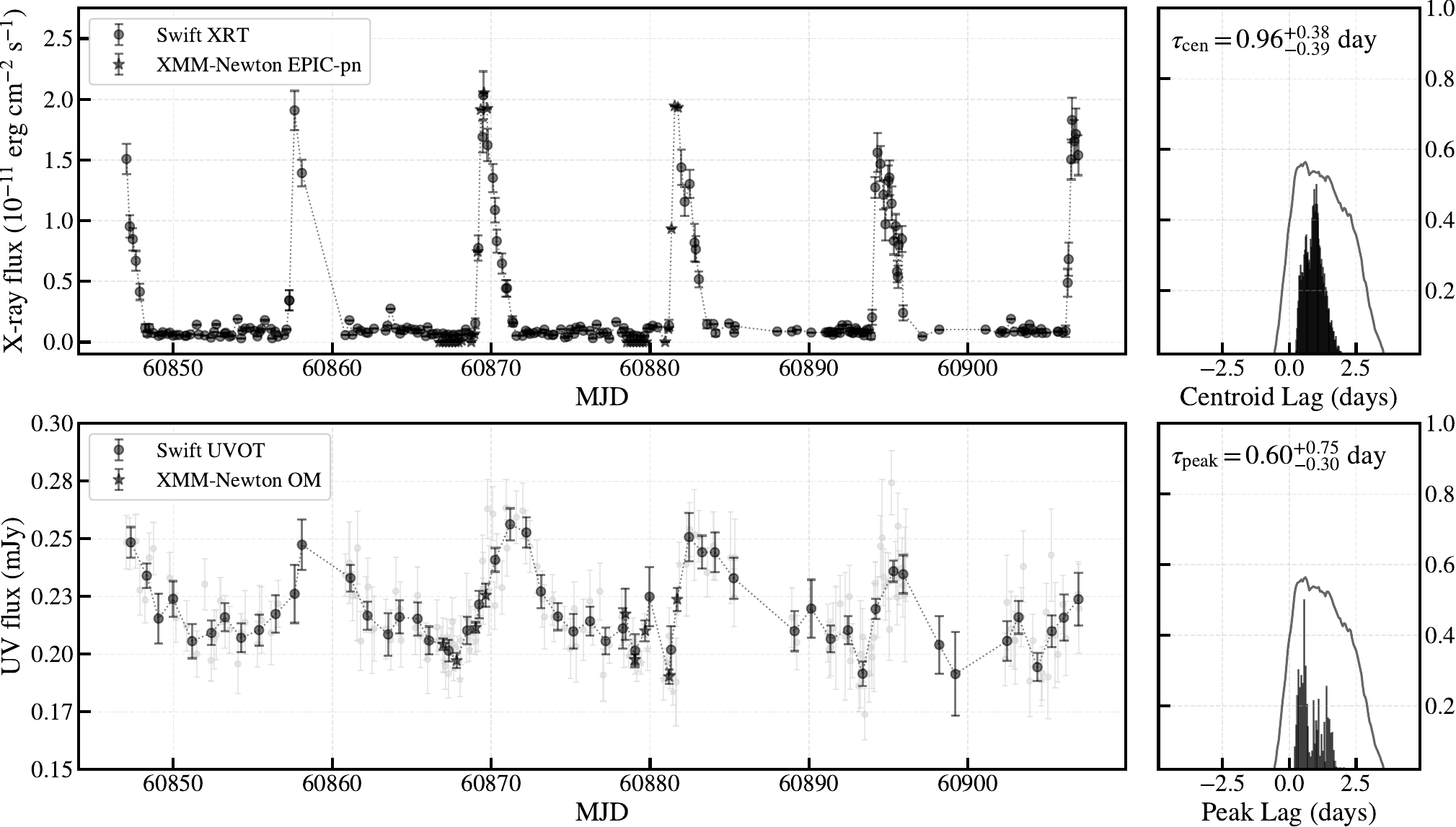}
\caption{Correlation between the soft X-ray (0.3--2~keV) and UV light curves of Ansky. Left panels: The X-ray light curves are obtained from {\it Swift} XRT and {\it XMM-Newton} EPIC-pn, while the UVW2 light curves are from {\it Swift} UVOT and {\it XMM-Newton} OM. The \textit{XMM-Newton} data are binned to 0.2~days to match the {\it Swift} cadence (original data shown in Fig.~\ref{fig:xmm}). The original {\it Swift} UV data and the 0.2-day-binned \textit{XMM-Newton} data are used for lag measurements (light grey); both are additionally binned to 1~day for clarity. Right panels: Cross-correlation results show the centroid lag (top), peak-lag distribution (bottom), and the corresponding CCF and lag posterior distributions. The maximum correlation coefficient is \textbf{$r_{\rm max}=0.56\pm0.04$}.
}
\label{fig:lag}
\end{figure*}

\section{Data Analysis} \label{sec:obs}

\subsection{Swift XRT and UVOT}

We use the Swift X-ray dataset compiled by \citet{Chakraborty26}, which includes all observations processed through February 2026. The X-ray Telescope (XRT) data were retrieved via the online interface\footnote{\url{https://www.swift.ac.uk/LSXPS}} to the Living Swift XRT Point Source (LSXPS) catalogue \citep{Evans23}, an automatically updated repository of Swift XRT observations obtained in Photon Counting (PC) mode with exposures longer than 100~s. The LSXPS provides count rates (counts s$^{-1}$), which we converted to flux units adopting a conversion factor of 1~cps $=2.2\times10^{-11}\ \mathrm{erg\ cm^{-2}\ s^{-1}}$. This factor was derived with WebPIMMS assuming a 100 eV blackbody spectrum at a redshift of 0.024 and Galactic absorption with $N_{\rm H}=2.6\times10^{20}\ \mathrm{cm^{-2}}$ \citep{Hernandez-Garcia25a}.

We performed aperture photometry for each $Swift$/UVOT filter using the \texttt{uvotmaghist} tool, employing a source radius of 5$\arcsec$ and a background radius of 20$\arcsec$ located in a source-free region. 
Galactic extinction was corrected using the law from \citet{Gordon2023ApJ} with $R_\mathrm{V}=3.1$ and $E(B-V)=0.0288$ \citep{Zhu25}. The near ultraviolet (NUV, $\lambda_{\rm eff}\sim 2310\rm{\AA}$) and far ultraviolet (FUV, $\lambda_{\rm eff}\sim1528\rm{\AA}$) magnitudes of the host galaxy are $18.89\pm0.07$ and $19.86\pm0.16$ measured from GALEX observations \citep{Osborne2023ApJS}. The extinction-corrected UV fluxes are interpolated or extrapolated to the three UV filters of UVOT by assuming a power-law spectrum of the two GALEX bands. The long-term UV light curves with the host galaxy subtraction are presented in Fig.~\ref{fig:longterm}, along with ZTF $g$- and $r$-band difference-image light curves from the ZTF Forced Photometry Service \citep{Masci2023arXiv}. We excluded the $Swift$/UVOT results during August 2023 to April 2024 due to the degraded spacecraft attitude control \citep{cenko2024GCN,Hernandez-Garcia25a}, and the results extracted in the low sensitivity areas.

\subsection{XMM EPIC-pn and OM}

We obtained four XMM-Newton pointed observations through GO program 096454 in AO 24 (PI: Chakraborty). The observations were taken on July 10/12/22/24, 2025 (OBSIDs 0964540101-0964540401); the first and third observations captured the quiescence, and the second and fourth captured the rise-to-peak of two consecutive eruptions. The data were reduced using XMM SASv21.0.0 and HEASoft v6.33. 

EPIC-pn source products were extracted from a circular region of 33$''$ radius and the background was extracted from source-free circular region on the same detector with a 60$''$ radius. We kept events with \texttt{PATTERN}$\leq 4$ (singles and doubles) and discarded time intervals with a 10–12 keV count rate $\geq 1$ counts s$^{-1}$ to mitigate background contamination. X-ray event files were extracted with \texttt{evselect}, then corrected for detector efficiency, vignetting, PSF, and bad pixels, and binned to 1~ks in the 0.3-2 keV band via the \texttt{epiclccorr} command. Optical monitor (OM) data were taken in FAST mode with the UVW2 filter and reduced using the \texttt{omfchain} command with a 6$''$ source extraction region and a \texttt{timebinsize} $=1100$~seconds to evenly divide the 4400 second exposures. The final light curves are shown in Fig. \ref{fig:xmm}.

\subsection{Lag Measurements and Correlation Reliability}
To quantify the time delay between the X-ray and UV variability in Ansky, we employ the interpolated cross-correlation function (\texttt{ICCF}; \citealt{Gaskell87,White94}), a traditional and widely adopted method that measures inter-band time lags based purely on the correlation between light curves. It does not rely on assumptions about a specific variability model or transfer function, making it a straightforward and minimally model-dependent estimator. In practice, we use the \texttt{PyI$^2$CCF} tool\footnote{\url{https://github.com/legolason/PyIICCF}} \citep{Guo21}, which not only measures lags but also evaluates the reliability of the cross-correlation through Monte Carlo simulations. We adopt a lag search window of $-5$ to $+5$ days with a sampling interval of 0.05 day. The time delay is characterized using both the peak lag ($\tau_{\rm peak}$), defined as the location of the maximum cross-correlation coefficient, and the centroid lag ($\tau_{\rm cen}$), computed from correlation coefficients above 80\% of the peak value. Uncertainties are estimated using the standard flux randomization and random subset selection (FR/RSS) method \citep{Peterson98} with 10000 realizations.

To assess the reliability of the correlation, we generate artificial light curves based on a damped random walk (DRW) model \citep[e.g.,][]{Kelly09} that match the sampling and signal-to-noise properties of the data, and perform null-hypothesis testing to compute the probability (p-value) that uncorrelated red-noise light curves could produce a correlation as strong as the observed one. Smaller p-values indicate more robust detections of correlated variability. Further details are provided in Appendix \ref{app:iccf} and \citet{U22}.

\section{Delayed UV counterpart to X-ray QPE}\label{sec:res}

Long-term monitoring of Ansky in UV/optical bands is shown in Fig.~\ref{fig:longterm}. The system first exhibited a major optical nuclear flare around MJD~$\sim59000$, followed by a gradual and coherent decline in both the UV and optical bands, accompanied by significant UV variability in timescales of years \citep{Hernandez-Garcia25a}. While X-ray emission was not detected until MJD~$\sim60350$, approximately 3.5 years after the optical peak—resembling the delayed X-ray emergence observed in some TDEs \citep{Gezari21}, although the underlying physical origin may be different. Since mid-2024, Ansky has exhibited a steadily lengthening X-ray recurrence period, increasing from $\sim4.5$ days to $\sim10$ days in mid-2025 and reaching $\sim14$ days by early 2026. This roughly stable evolution, characterized by $\dot{P} \equiv \Delta P / \Delta t \sim 0.017$ day day$^{-1}$ and a potential additional $\sim$150-day modulation \citep{Hernandez-Garcia25b,Chakraborty26}, provides a stringent observational constraint for future theoretical models.

The high-cadence {\it Swift} multi-wavelength campaign during MJD~60840--60910 in Fig.~\ref{fig:lag} captured approximately five consecutive X-ray QPE cycles. Remarkably, the light curve in the UVW2 band displays recurrent brightening events that are temporally coincident with the X-ray QPEs. These UV flares appear slightly broader and less impulsive than the sharp, high-amplitude spikes seen in X-rays, with widths roughly twice as large based on visual inspection. Moreover, the UV flare profile appears to begin slightly earlier than the X-ray flare, resembling the behavior observed between different X-ray energy bands in eRO-QPE1 \citep{Arcodia22}. However, given the very different detection limits of the two bands, this offset should be interpreted with caution. The directly observed UV flux shows a modest variability of $\sim$30\% and remains below a factor of two even after accounting for the host contribution and Galactic extinction, whereas the X-ray variability is over 500 times stronger (see also Appendix \ref{app:xmm}).

The nearly synchronous variations reveal a moderate correlation between the X-ray and UV light curves, yielding a maximum correlation coefficient of \textbf{$r_{\rm max}=0.56\,\pm\,0.04$} based on the original {\it Swift} and 0.2-day binned {\it XMM-Newton} data\footnote{The \textit{XMM-Newton} light curves are binned to match the Swift cadence of 0.22 days, in order to prevent the more densely sampled \textit{XMM-Newton} data from disproportionately weighting the lag measurement.}. The null-hypothesis test gives a p-value of 6$\times$10$^{-4}$ indicating that only 6 of the 10,000 simulated uncorrelated light-curve pairs produced a correlation as strong as the observed one. This means that, although the correlation coefficient itself is moderate, primarily due to incomplete sampling of the flare peaks, the correlation is statistically robust. The UV variations systematically lag the X-ray eruptions, with a centroid lag of $0.96^{+0.38}_{-0.39}$~days and a peak lag of $0.6^{+0.75}_{-0.30}$~days in the observed frame. Given the small redshift, the difference between observed- and rest-frame lags is negligible compared to the uncertainties, so we do not distinguish between them hereafter. We adopt the $\tau_{\rm cen}$ as the fiducial value because it provides a weighted average over the high-correlation region of the CCF and is less sensitive to noise and asymmetric response functions than the peak lag. 

Therefore, the delayed UV response, together with its substantially reduced variability amplitude and broader temporal profile, strongly suggests that the X-rays more directly trace the primary QPE emission, whereas the UV is more likely a secondary by-product.

\begin{figure*}[ht!]
\includegraphics[width=1.0\linewidth]{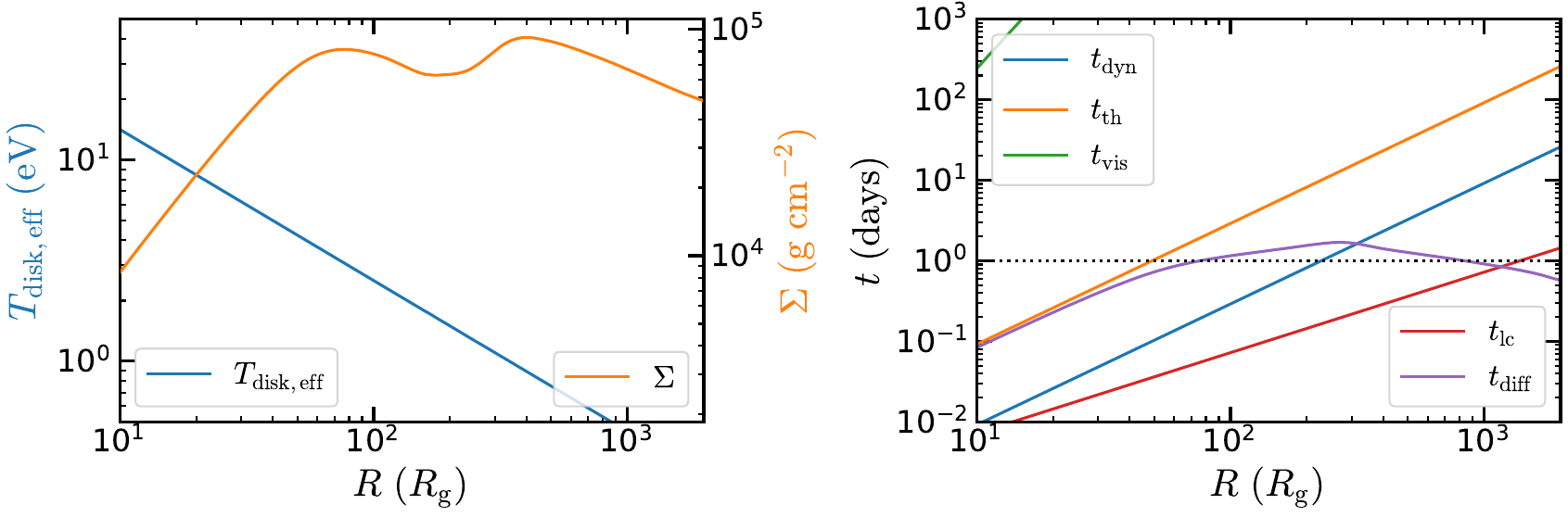}
\vspace{-20pt}
\caption{Disk profiles (left panel) and different related timescales for the thin disk model (right panel). The blue line in the left panel shows the disk effective temperature ($T_{\rm disk,eff}$) and the orange line represents the disk surface density $\Sigma$ with an SMBH mass of $M_{\rm BH}=5\times10^{6}M_{\odot}$, disk viscosity $\alpha=0.1$, and accretion rate $\dot{m}=0.03$. Lines of different colors in the right panel represent the dynamical, thermal, viscous, light-crossing, and diffusion timescales calculated using the same set of parameters. The observed delay time $\sim1$ days is shown as the dotted line in the bottom panel. 
}
\label{fig:disk_delay}
\end{figure*}

\section{Discussions }\label{sec:dis}
\subsection{Why Is the UV Counterpart Detectable in Ansky?}
Tab.~\ref{tab:tab2} summarizes the currently established QPE sample, including their recurrence times, burst durations, peak luminosities, estimated black hole masses (subject to substantial uncertainties), and UV properties. Among these sources, Ansky stands out for having the longest recurrence period and burst duration. It also has the largest amplitude (peak-to-quiescence ratio) and the highest peak luminosity in the present sample, although its peak luminosity does not differ significantly from that of other luminous QPEs.

A comparison with other QPEs suggests that three factors may enable the detection of UV variability response to X-ray QPE in Ansky. First, because $T_{\rm rec}$ significantly exceeds the measured X-ray–UV lag, the UV response to each individual flare remains temporally separated rather than blended with adjacent events. In systems with shorter recurrence intervals, responses to successive flares may overlap, smoothing the variability and reducing the visibility of periodic modulation. Second, the large flare amplitude enhances the UV contrast above the quiescent level, probably also leading to relatively longer apparent burst durations and making the variability easier to detect. Finally, for previously known QPE sources with relatively long recurrence times ($\gtrsim$ 1 day), sufficiently high-cadence UV monitoring over multiple consecutive cycles is generally lacking, and any corresponding UV variability, even if present, may therefore remain undetected.

\subsection{Origin of the X-ray-UV lag}\label{sec:lag}

The measured soft X-ray–UV lag in Ansky is reminiscent of similar delays reported in nearby broad-line active galactic nuclei (AGNs). In the lowest-mass system, NGC~4395 ($\log M_{\rm BH}/M_{\odot}\sim4$–5), the lag is only $\sim8$ minutes \citep{McHardy16}. At intermediate masses of $\log M_{\rm BH}/M_{\odot}\sim7$, sources such as NGC~4593 and NGC~7469 exhibit delays of $0.66\pm0.15$ days \citep{McHardy18} and $0.72\pm0.51$ days \citep{Pahari20}, respectively. More massive systems with $\log M_{\rm BH}/M_{\odot}\sim7.5$–8.5, including NGC~2617 \citep{Shappee14}, NGC~4151 \citep{Edelson17}, NGC~5548 \citep{Edelson15}, and Fairall~9 \citep{HS20}, show longer lags of order $\sim$1–3 days. Given a measured lag of $\lesssim1$ days, Ansky closely matches the delays observed in intermediate-mass AGNs such as NGC~4593 and NGC~7469, although its black hole mass remains uncertain ($\log M_{\rm BH}/M_{\odot}\sim6$–7 from indirect estimates). The similarity of its lag to those seen in nearby AGNs suggests that a comparable physical origin may underlie the variability, despite the absence of hard X-ray and prominent broad optical emission lines in QPEs.

To investigate the physical origin of the observed $\sim$1~day lag between the soft X-ray (0.5~keV) and UVW2 ($\sim$2000\AA) emissions, we compare the relevant timescales for a thin disk model \citep{Sirko03}.  
To make a quantitative comparison, we first calculate the disk structures by assuming an $\alpha$-viscosity with $\alpha=0.1$ \citep{ShakuraSunyaev73}, and a disk accretion rate of $\dot{m}=0.03$ (in unit of Eddington accretion rate; \citealt{Chakraborty26}), and an SMBH with $M_{\rm BH}=5\times10^{6}M_{\odot}$. The radial profiles of disk temperatures and surface density are shown in the left panel of Fig.~\ref{fig:disk_delay}. 
The X-ray QPE emission should come from some localized heated region (e.g., shock) instead of the disk itself due to the low disk temperature.
However, the disk itself could contribute to the UV emission around $10^{3}~R_{\rm g}$, where $R_{\rm g}$ is the gravitational radius of the SMBH.

Now we discuss the origin of the X-ray-UV lag. 
We first show that the viscous timescale $t_{\rm vis}\approx t_{\rm th}(R/H)^{2}$ ($t_{\rm th}$ is the disk thermal timescale) in Fig.~\ref{fig:disk_delay},  is much longer than the observed time delay, excluding radial mass transport as the origin of the correlated variability.
While the thermal timescale $t_{\rm th}$ is numerically comparable to the observed $\sim1$-day lag for $R\lesssim100\ R_{\rm g}$, it is unlikely to account for the coherent nature of the time lag between the X-ray QPE and its UV counterpart.

Following empirical results from continuum reverberation mapping (CRM; e.g., \citealt{Fausnaugh16}), we estimate the light-crossing timescale as $t_{\rm lc}=X R_{\rm UV}/c$ with a correction factor $X=2.5$, shown as a black line in the right panel of Fig.~\ref{fig:disk_delay} using the same model parameters adopted above, where $R_{\rm UV}$ is the size of the UV emission region. We can see that this timescale is roughly consistent with the observed delay time of $\sim1$ days when the UV emission arises from $R\sim10^{3}\ R_{\rm g}$. 
The larger radial extent of the AGN disk, relative to the compact accretion disk formed following a TDE, may provide another explanation for the detection of UV counterparts to X-ray QPEs. In this scenario, the outer region of the disk reprocesses the X-ray emission from the inner region (e.g., in the disk instability and mass transfer models), which contributes to the X-ray--UV lag. However, this X-ray reprocessing scenario may have difficulty explaining the earlier rise of the UV flare relative to the X-ray one, if confirmed by future observations.

Another related timescale is the diffusion timescale $t_{\rm diff}$, which can be estimated as a modified reprocessing timescale $t_{\rm diff}\simeq \tau R_{\rm X}/c$. Here, $\tau$ is the optical depth, and $R_{\rm X}$ is the size of the X-ray emission region which is estimated to be $R_{\rm X}\simeq 1.5\times10^{11}\ {\rm cm}$ \citep{Hernandez-Garcia25a}. 
We adopt the disk optical depth as an approximation of the optical depth of the X-ray-emitting region. In this scenario—most relevant to the star-disk collision model—both the UV and X-ray emission originate from the same expanding shock-heated region. The observed time delay is then simply an optical depth effect: the higher-energy X-rays peak first, followed by a delayed UV peak. We show the numerical results as the magenta lines in the right panel of Fig.~\ref{fig:disk_delay}, which is comparable to the observed delay time.

Moreover, the diffusion timescale for an expanding blob can also be approximated by Eq. 13 of \cite{Vurm25}, which reads as $t_{\rm diff} \approx \bigg(\frac{\kappa_{\rm T} M_{\rm ej}}{4\pi cv_{\rm ej}}\bigg)^{1/2}$, where $M_{\rm ej}$ is the ejected disk mass heated by shock, $\kappa_{\rm T}$ is the Thomson scattering opacity, and $v_{\rm ej}$ is the ejecta speed. Using the disk profile shown above, we can obtain a diffusion timescale $\sim1$ day, which is also roughly consistent with the observed time lag.

\subsection{Constraints on the QPE models}

Several mechanisms have been proposed to explain the X-ray QPE to date. For Ansky, however, any viable model should account for three key observational facts: (1) the detection of a UV counterpart to the X-ray QPE; (2) a UV delay of order one day relative to the X-rays; and (3) a steadily increasing QPE recurrence period \citep{Chakraborty26}. In the following, we briefly discuss the new constraints these observations impose on current QPE models.

In the disk-star collision scenario (where ``star'' may also refer to a stellar-mass black hole or an extended trail of bound stellar debris), the X-ray QPEs could arise from a compact hot spot formed by shock-heated gas generated during the collision with the TDE-formed disk. The expansion of the hot spot leads to delayed UV emission relative to the X-ray as the photons undergo adiabatic losses, appearing over the diffusion timescale we have discussed in Section~\ref{sec:lag}. 
From the disk surface density profile $\Sigma\propto R^{3/2}\dot{m}^{-1}$ as shown in Fig.~\ref{fig:disk_delay}, it is apparent that the long recurrence times and low quiescence luminosity imply a large disk density, implying efficient thermalization of shocked disk material, rather than the ``photon-starvation'' regime assumed for shorter-period QPEs \citep{Linial23b}.
Efficient thermalization will have an important effect on the observed emission band, as with greater photon production the observed emission will appear softer \citep{Vurm25}. 
The high disk density also implies large ejecta masses, which will increase the timescale for photosphere retreat result in greater adiabatic losses of the escaping photons \citep{Chakraborty25b}.
Actually, using the diffusion timescale and the assumed ejecta velocity of $0.03c$, the ejecta size can be estimated as
$R_{\rm diff} \simeq v_{\rm ej} t_{\rm diff} \simeq 7 \times 10^{13}\ {\rm cm}$,
which implies a UV flux of $\sim 0.1\ {\rm mJy}$ from the expanding blob, in reasonable agreement with the observed UV counterpart.

However, the disk-star collision model faces a key challenge in explaining the time evolution of the QPE recurrence period. One possible mechanism for producing an increasing recurrence time is mass transfer from the star to the SMBH \citep[e.g.,][]{Sepinsky09,Dosopoulou16b}, though this interpretation remains highly uncertain \citep{Chakraborty26}. An alternative possibility has emerged from recent work suggesting that an \textit{accreting} stellar-mass object embedded in an accretion disk may undergo orbital expansion due to the influence of a mini-disk formed around it \citep{Li24,Laune24,Pan26}. However, whether this mechanism can quantitatively account for the period ratio change observed in Ansky remains an open question.

For the periodic mass transfer model, the observed time lag could result from reprocessing of inner-disk X-ray emission by the outer disk when the outer disk becomes warped or flared --- a delay that can be approximated by the light-crossing timescale as discussed in Section~\ref{sec:lag}. In this scenario, the increase in recurrence time could be attributed either to mass transfer from a lower-mass companion to the SMBH \citep[e.g.,][]{Sepinsky09,Dosopoulou16b} or to general relativistic precession \citep[e.g.,][]{Franchini23,Linial24,Zhang25,Deng25}. However, both explanations also face significant challenges \citep{Chakraborty26}.

The disk instability model offers an alternative scenario. In this framework, the delayed UV emission could arise from the similar outer disk reprocessing of X-rays from the inner unstable region as mentioned above. A steady few-percent increase in accretion rate has been proposed to explain the evolution of the QPE recurrence time \citep{Pan25}. However, this model faces at least two challenges. The physical mechanism capable of driving such a stable, gradual increase in accretion rate remains unclear. Additionally, the model struggles to reproduce the characteristic fast-rise and slow-decay shape of the QPE light curve, as well as the energy-dependent peak delay. While it has been suggested that these discrepancies could be mitigated if the unstable region propagates outward \citep{Pan22}, a quantitative comparison with observations is still lacking.

In summary, the time delay we observe between the UV and X-ray QPEs can be interpreted as the diffusion timescale of an expanding hot blob within the star–disk collision framework. Nevertheless, we cannot exclude the possibility that the lag arises from the light-crossing timescale associated with X-ray reprocessing in the outer AGN disk, as proposed in disk instability or mass-transfer models.
More broadly, concerning the physical origin of the QPE emission itself, all existing models encounter substantial difficulties in simultaneously explaining both the UV lag and the evolution of the recurrence time observed in Ansky. This tension highlights the need for more detailed theoretical investigations.

\begin{table*}[h]
\centering
\caption{Known X-ray QPE Sources}
\label{tab:tab2}
\begin{tabular}{lccccccr}
\hline
Name & Redshift & $T_{\rm rec}$ & $T_{\rm dur}$ & $\log L_{\rm peak}$ (erg s$^{-1}$) & $\log M_{\rm BH}/M_\odot$ & UV Properties & Ref. \\
\hline
Ansky & 0.024 & 4.5$\rightarrow$14 d & 0.6$\rightarrow$1.5 d & 43.6 & $6.34\pm0.66$ & significant var. & 1, 2, 3 \\
AT2019qiz & 0.015 & 39--57 h & 8--10 h & 43.2 & $6.27 \pm 0.76$ & potential var. & 4 \\
AT2022upj & 0.054 & 0.5--3.5 d & 0.3--1 d & 43.0 & $6.38 \pm 0.56$ & no var. & 5 \\
eRASSt J2344 & 0.100 & 12 h & 2 h & 43.4 & $7.20 \pm 0.20$ & no var. & 6 \\
eRO-QPE1 & 0.051 & 18.5 h & 8 h & 43.3$\rightarrow$41.6 & $5.90 \pm 0.79$ & no var. & 7, 8 \\
eRO-QPE2 & 0.018 & 2.4$\rightarrow$2.3 h & 0.5 h & 42.0 & $5.43 \pm 0.79$ & no var. & 7, 9 \\
eRO-QPE3 & 0.024 & 20.4 h & 2--2.5 h & 42.6$\rightarrow$41.4 & $5.53 \pm 0.79$ & no var. & 10 \\
eRO-QPE4 & 0.044 & 9.8--14.7 h & 1.1 h & 43.1 & $7.31 \pm 0.75$ & no var. & 10 \\
eRO-QPE5 & 0.116 & 3.7 d & 0.6 d & 43.0 & $7.45 \pm 0.52$ & no var. & 11 \\
GSN 069 & 0.018 & 8--10 h & 1.3 h & 42.7 & $6.28 \pm 0.72$ & no var. & 12, 13, 14 \\
RX J1301 & 0.023 & 3.6--5.6 h & 0.3 h & 42.1 & $6.14 \pm 0.88$ & no var. & 15, 16 \\
XMM J0249 & 0.019 & 2.5 h & 0.3 h & 41.5 & $5.29 \pm 0.55$ & corresponding dip & 17 \\
\hline
\end{tabular}
\tablecomments{Peak X-ray luminosities ($L_{\rm peak}$) is generally below 2 keV. Black hole masses are averaged values from \citet{Wevers22,Wevers24,Wevers25,Arcodia25}. Apart from Ansky, UV variability is generally absent, with only AT2019qiz showing a possible weak UV variation and XMM J0249 exhibiting a UV dip coincident with an X-ray eruption. The arrow indicates a relatively clear evolutionary trend. AT2019vcb \citep{Quintin23} is not included here, as further evidence is required to confirm its QPE nature.
References: 1. \citet{Sanchez-Saez24}; 2. \citet{Hernandez-Garcia25a}; 3. \citet{Hernandez-Garcia25b}  4. \citet{Nicholl24}; 5. \citet{Chakraborty25}; 6. \citet{Baldini26}} ; 7. \citet{Arcodia21}; 8. \citet{Chakraborty24} 9. \citet{Arcodia24b}; 10. \citet{Arcodia24a}; 11. \citet{Arcodia25}; 12. \citet{Miniutti19}; 13. \citet{Miniutti23a}; 14. \citet{Miniutti23b}  15. \citet{Sun13}; 16. \citet{Giustini20}; 17. \citet{Chakraborty21}.
\end{table*}

\setcounter{figure}{0}
\renewcommand{\thefigure}{S\arabic{figure}}
\begin{figure*}[htbp]
  \centering
   \includegraphics[width=\linewidth]{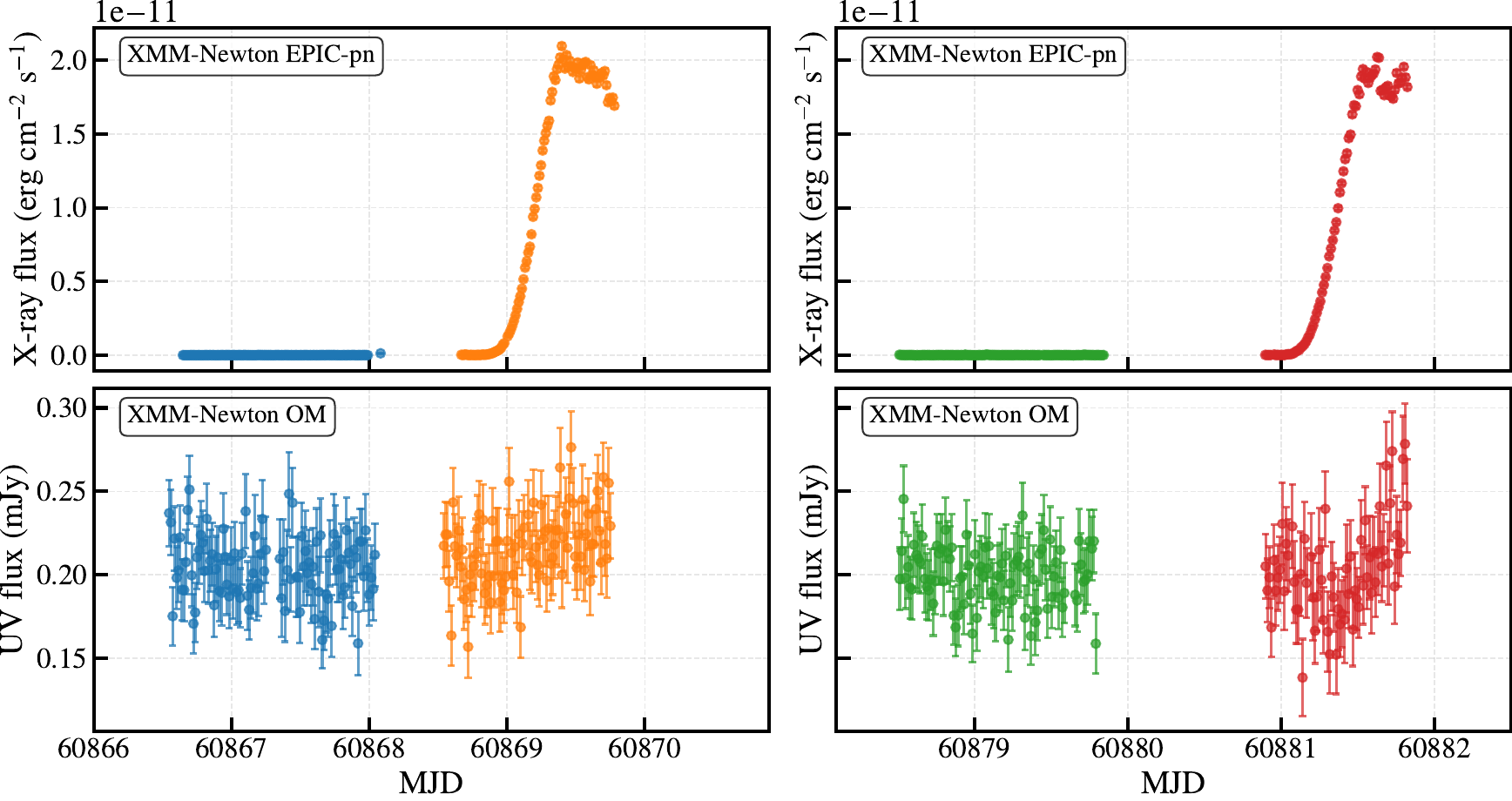}
\caption{XMM-Newton EPIC-pn (0.2-10 keV) X-ray (top) and OM UVW2 UV (bottom) light curves showing two flare episodes. Different colors correspond to four XMM observations (OBSIDs 064540101-0964540401).}
\label{fig:xmm}
\end{figure*}

\renewcommand{\thefigure}{S\arabic{figure}}
\begin{figure*}[htbp]
  \centering
   \includegraphics[width=\linewidth]{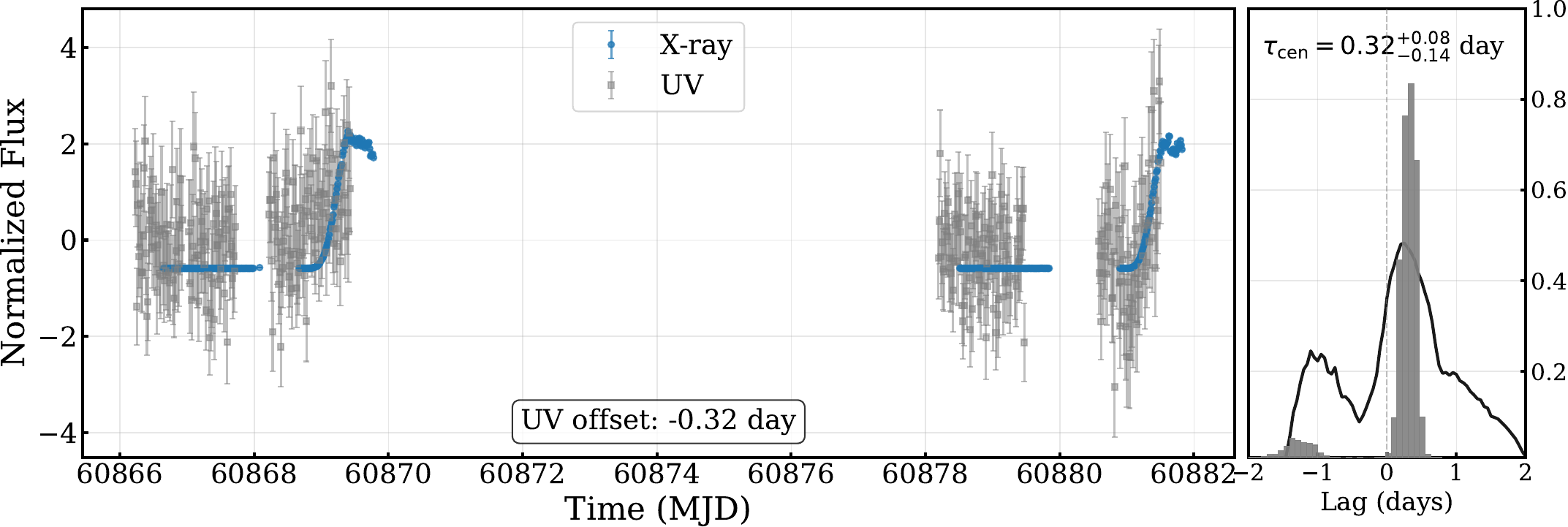}
\caption{The left panel shows the \textit{XMM-Newton} X-ray and UV light curves, with the UV data shifted by the measured centroid lag of 0.32 day, while the right panel shows the lag posterior distribution and the corresponding CCF. X-ray and UV light curves are normalized to zero mean and unit variance; errors are scaled accordingly.}
\label{fig:xmm_ccf}
\end{figure*}

\begin{acknowledgments}
We thank the anonymous referee for the constructive comments that helped improve this paper. This work is supported by the National Key R\&D Program of China (No.~2023YFA1607903). HXG acknowledges support from the National Natural Science Foundation of China (NSFC; Nos.~12473018 and 12522304) and the Overseas Center Platform Projects of CAS (No.~178GJHZ2023184MI). ZY acknowledges support from the NSFC (Nos.~12373049 and 12361131579). YPL acknowledges support from the NSFC (Nos.~12373070 and 12192223) and the Natural Science Foundation of Shanghai (No.~23ZR1473700). The calculations have made use of the High Performance Computing Resource in the Core Facility for Advanced Research Computing at Shanghai Astronomical Observatory. LHG acknowledges financial support from ANID program FONDECYT Iniciaci\'on 11241477. MFG is supported by the NSFC 12473019, the Shanghai Pilot Program for Basic Research-Chinese Academy of Science, Shanghai Branch (JCYJ-SHFY-2021-013), the National SKA Program of China (Grant No. 2022SKA0120102), and the China Manned Space Project with No. CMS-CSST-2025-A07. SLL is supported by NSFC 12273089. WWZ is supported by the Shanghai Municipal Natural Science Foundation Youth Project of the 2025 Basic Research Program Grant No. 25ZR1402546. PA acknowledges support from ANID via FONDECYT Regular 1241422, Millennium Science Initiative Program NCN$2023\_002$ and CIDI N. 21 U. de Valpara\'iso, Chile

\end{acknowledgments}





%
\facilities{Swift (XRT and UVOT), XMM (EPIC-pn and OM)}

\software{astropy \citep{2013A&A...558A..33A,2018AJ....156..123A,2022ApJ...935..167A} }


\appendix 

\section{The long-exposure {\it XMM-Newton} monitoring of Ansky}
\label{app:xmm}
Coordinated long-exposure {\it XMM-Newton} monitoring in July 2025, shown in Fig.~\ref{fig:xmm}, captured two extreme X-ray flares; however, the coverage spans only the quiescent and rising phases of the corresponding UV flare. The turnover is clearly detected in the X-ray light curve but not in the UV owing to the intrinsic delay. During the quiescent state, the X-ray flux is $\sim2\times10^{-14}$~erg~cm$^{-2}$~s$^{-1}$ \citep{Hernandez-Garcia25a}, while the extinction-corrected ($\times$1.2), host-subtracted UV flux (0.09 $\pm$ 0.01 mJy) remains at $\sim0.12$~mJy; this residual emission may originate from a nascent accretion disk formed by a prior TDE \citep{vanVelzen19}. During the flaring state, the X-ray flux increases dramatically to $\sim2\times10^{-11}$~erg~cm$^{-2}$~s$^{-1}$, with an amplitude $\sim1000$ relative to the quiescent level, in contrast to the much more modest UV variations ($<2$ in flux change), even after accounting for host-galaxy contribution and Galactic extinction. Notably, a potential dip appears during the rising phase of the second flare; however, the underlying physical origin remains unclear.

In addition, we explore the potential lag using the \textit{XMM-Newton} data alone. Using the current binning strategy (both X-ray and UV data binned to $\sim$1~ks), we obtain an ICCF lag of $0.32^{+0.08}_{-0.14}$ days (Fig.~\ref{fig:xmm_ccf}). However, an important caveat is that the X-ray light curve captures the flare peak, whereas the UV light curve does not. Consequently, during the CCF calculation the UV light curve can overlap with {\it arbitrary} rising portions of the X-ray flare, which is also reflected in the lag posterior distribution spanning $\sim0-0.4$ day. In the left panel, where the light curves are shifted by the median centroid lag, the UV data points align closely with the X-ray turning point. This positioning implies that the inferred lag of $\sim$ $0.2-0.4$ day should be regarded as a lower limit rather than a robust lag detection. This also motivates our adoption of the 0.2 day binning in Fig. \ref{fig:lag}, which provides a sampling comparable to that of the \textit{Swift} data for the joint lag measurement. Future observations with denser and more continuous X-ray/UV coverage, particularly those capturing multiple flare peaks in both bands, will be crucial for resolving this ambiguity and establishing a more robust lag measurement.

\section{Significance of Cross correlation}
\label{app:iccf}

To evaluate the significance of the cross-correlation coefficient and the measured lag, we perform Monte Carlo simulations using a DRW model as the null hypothesis. Stochastic variability in accreting black hole systems is often described by DRW processes on timescales of days to years \citep[e.g.,][]{Kelly09,MacLeod10,Zu11}. We first fit the observed UV light curve with \texttt{PyI$^{2}$CCF} to determine the DRW parameters, including the damping timescale ($\tau$) and asymptotic variability amplitude ($\sigma$). These parameters are then used to generate 10,000 simulated UV light curves that preserve the same variability properties, cadence, and measurement uncertainties as the observations, while the X-ray light curve is kept fixed. For each realization, we compute the ICCF between the simulated UV and observed X-ray light curves and record the maximum correlation coefficient $r_{\rm max}$. To avoid possible biases, a two-way simulation is adopted, with the CCF computed using real and simulated light curves in alternating roles. We also verify that both the X-ray and UV light curves of Ansky can be reasonably described by the DRW model, although their variability differs from those of typical AGN light curves.

This produces the expected distribution of $r_{\rm max}$ values for uncorrelated red-noise light curves. Among the 10,000 simulations, only 6 cases yield $r_{\rm max}$ values larger than the observed one ($r_{\rm max}=0.56\pm0.04$, with the uncertainty taken into account). This indicates that the observed UV--X-ray correlation is highly unlikely to arise from stochastic variability alone. In addition, we examine the centroid lags from the same 10,000 realizations and find that none of them produces negative lags, suggesting that the positive lag is statistically robust and unlikely to arise from random fluctuations. 



\bibliography{sample701}{}
\bibliographystyle{aasjournalv7}



\end{document}